\documentclass[12pt,onecolumn]{IEEEtran}
\usepackage{color}
\usepackage{amsfonts}
\usepackage{epsfig}
\usepackage{graphicx}
\usepackage{cite}
\usepackage{amsthm}
\usepackage{amsmath,amsfonts,amssymb,graphicx} %Èç¹û·½³ÌÖеÄʽ×ÓÐèÒª·ÖÐÐÏÔʾ£¨1a£©,(1b),(1c)µÄ»°¾ÍÓÃÕâ¸öÍ·Îļþ
\usepackage{setspace}

\begin{document}

\begin{spacing}{1.2}

\title{ Fractional-order Generalized Principle of Self-Support (FOG PSS)  in Control Systems Design}
 \author{Hua Chen$^{1^{*}, 3}$, and YangQuan Chen$^{2}$
 \thanks{$^1$  Mathematics and Physics Department, Hohai University, Changzhou Campus, Changzhou, 213022, China.}
 \thanks{$^2$ MESA Lab, University of California, Merced, 5200 North Lake Road, Merced, CA 95343, USA.}
  \thanks{$^3$ Changzhou Key Laboratory of Special Robot and Intelligent Technology, Changzhou, 213022, China.}
%  \thanks{$^4$ School of Science, Ningbo University of Technology, Ningbo,  315211, China.}
  \thanks{
 \texttt{\protect
 $^{*}$Corresponding author:Hua Chen(e-mail:chenhua112@163.com)}}
 }

\maketitle

\begin{abstract}
% with uncelebrated visual serving parameters,

This paper reviews research that studies the principle of self-support (PSS) in some control systems and proposes a fractional-order generalized  PSS framework for the first time. The existing PSS approach focuses on practical tracking problem of integer-order systems including robotic dynamics,
high precision linear motor system, multi-axis high precision positioning system with unmeasurable variables, imprecise sensor information, uncertain parameters
and external disturbances. More generally, by formulating the fractional PSS concept as a new generalized framework, we will focus in the possible fields on the fractional-order control problems such as practical tracking, $\lambda$-tracking, etc. of robot systems, multiple mobile agents, discrete dynamical systems, time delay systems and other uncertain nonlinear systems. Finally, the practical tracking of a first-order uncertain model of automobile is considered as a simple example to demonstrate the efficiency of the fractional-order  generalized principle of self-support (FOGPSS) control strategy.

\end{abstract}

\markboth{}{Shell \MakeLowercase{\textit{et al.}}: Bare Demo of
IEEEtran.cls for Journals}

% \fi
\IEEEpeerreviewmaketitle

\thispagestyle{empty}

\section{Introduction}

The conception of PSS can be described by the following crucial characteristics for the existence of each phenomenon [1]:
(1) \emph{Self-existence}, each phenomenon (thing, fact, single element, unit, set, system, process, ... ) is an entity with its own being and nature. It exists as something in (of, by) itself, not as any other thing. (2) \emph{Existence as a whole}, each phenomenon ( ... ) exists as a whole. It is, or has a wholeness
which includes all other phenomena. $`$Whatever comes into existence, always comes as a whole.' (Plato, The Sophist). (3) \emph{Existence in a whole}, no phenomenon ( ... ) exists entirely alone.
Each is a part of other phenomena. Indeed, observing Fig.1, in a recent report [2],
as Alley pointed out that the ice movement may affect the regional climate change and the changes in temperature so that the rising of the sea levels, but instead,
changes of the sea surface will also affect the ice movement, so they are reciprocally cause and effect,
they are interrelated and interact and constitute an integral whole (self-support as a whole).

Additionally, as seen in Fig.2,  the best representative example  for another \emph{self-referential} (see [1] and references therein) seems to be a medieval paradox, the
Uroboros the archetype of a \emph{vicious circle} formed by a snake, or a dragon, looped in a circle, biting its own tail.
How to distinguish which is the first which is the end, why would people do so: making clear which is cause and which is effect (Fig.3)? Based on the PSS idea, it
just a self complete whole- a self support system.

Then, as for control systems, how to consider it with these three \emph{existences} above with PSS?

\begin{center}
\includegraphics [scale=0.8,trim=200 30 200 10]{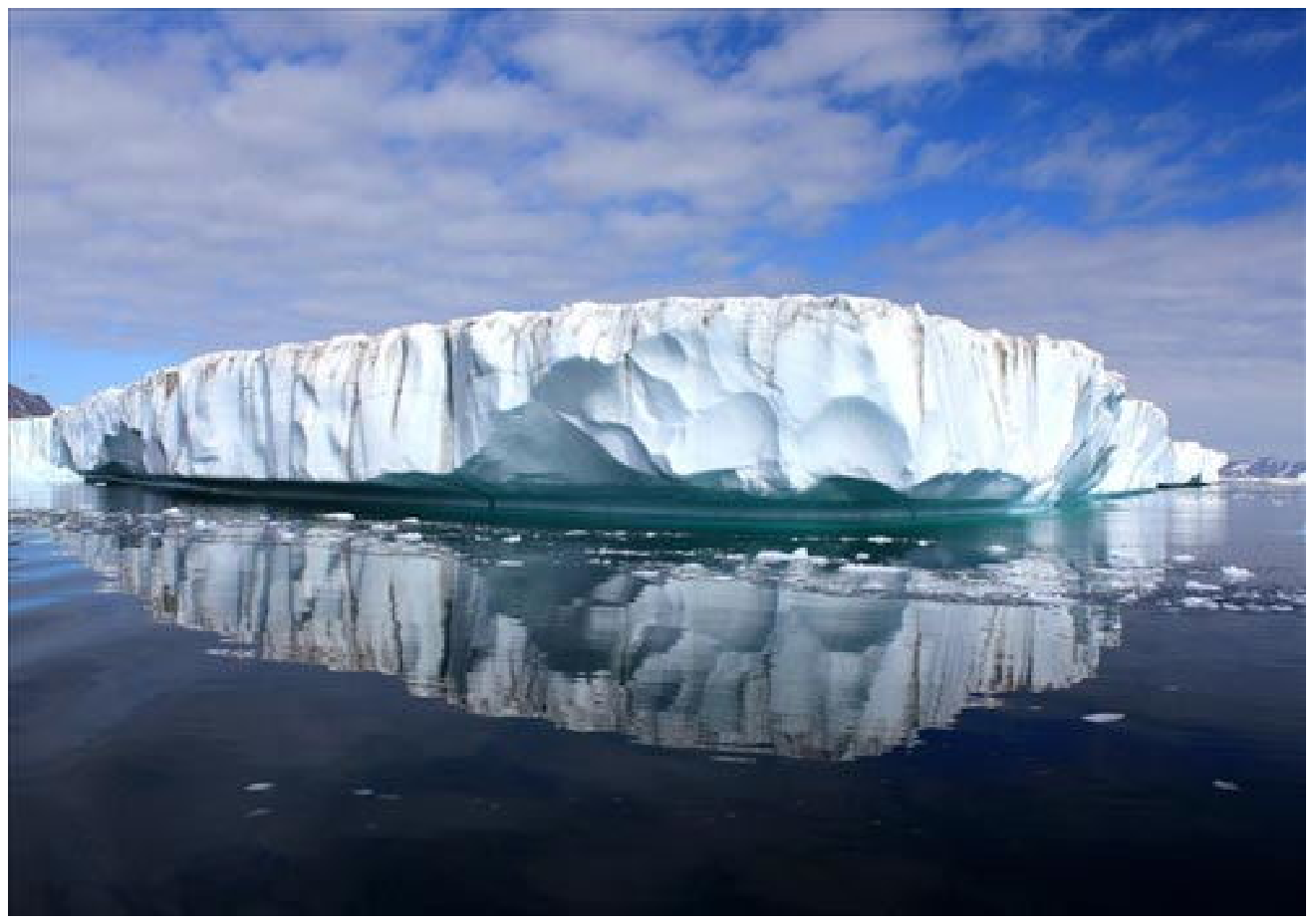}\\
\label{Fig4}{\fontsize{9.3pt}{11.6pt}\selectfont Fig.~1~~ Interaction between the ice movement and a rise in sea levels }
\end{center}

\begin{center}
\includegraphics [scale=0.8,trim=200 30 200 10]{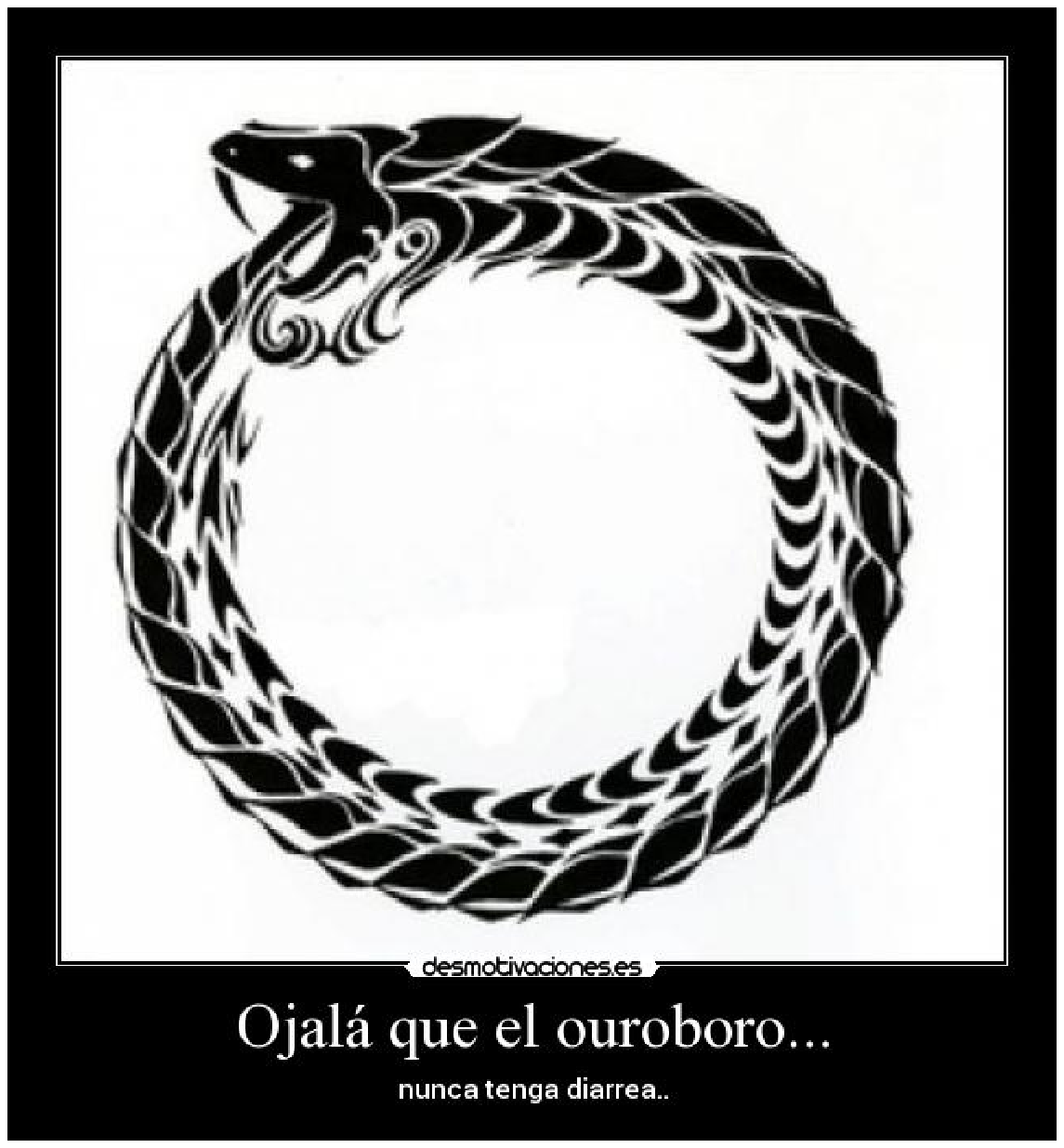}\\
\label{Fig4}{\fontsize{9.3pt}{11.6pt}\selectfont Fig.~2~~ One dragon/ Uroboros  }
\end{center}

\begin{center}
\includegraphics [scale=0.6,trim=200 30 200 10]{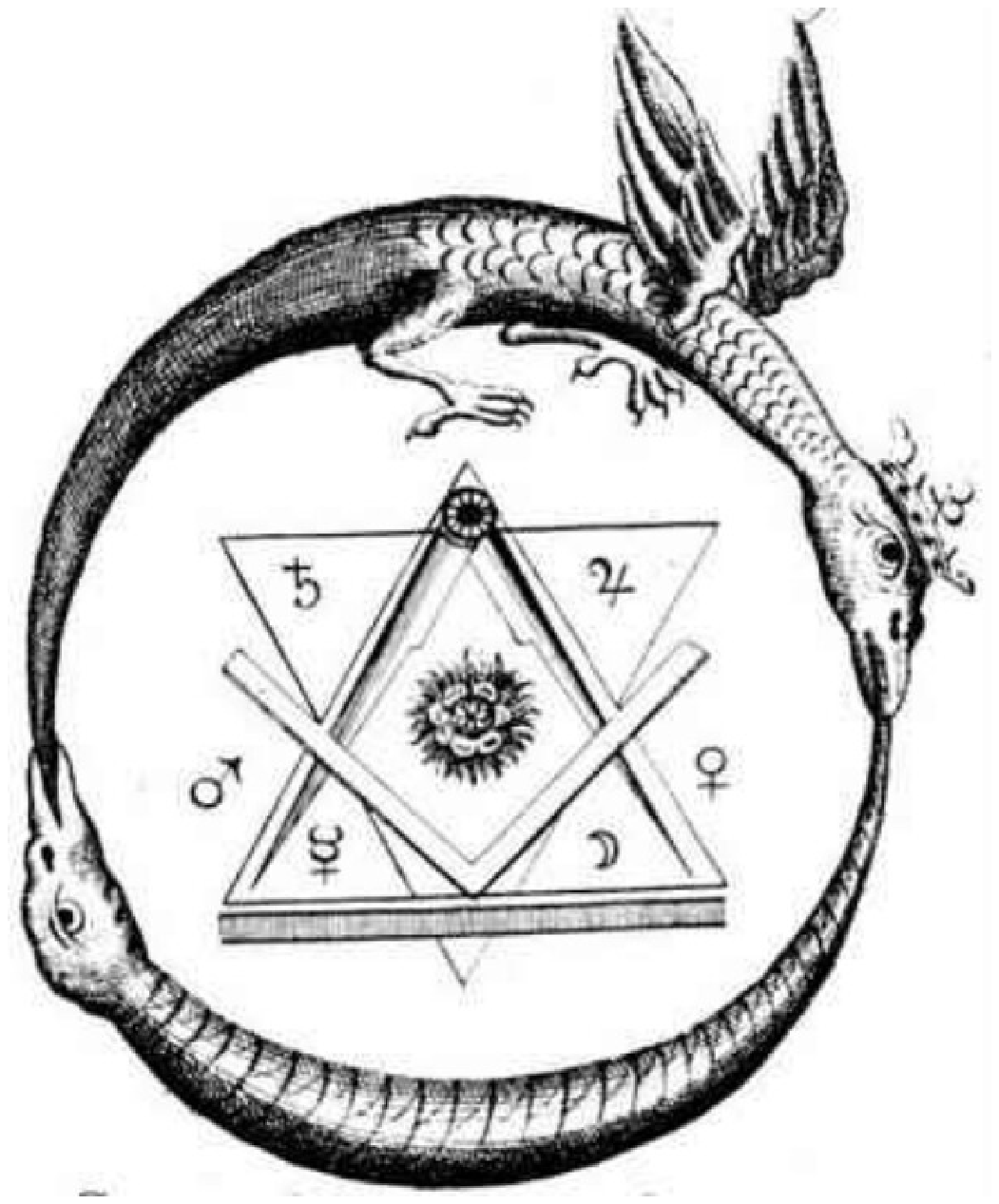}\\
\label{Fig4}{\fontsize{9.3pt}{11.6pt}\selectfont Fig.~3~~ Two dragons tail to mouth to tail}
\end{center}

A control signal (which is physically an amount of energy provided from the outside to a robotic system, in the form of either an input voltage or current
injected to the driving actuators) might be regarded as a self-supported variable, i.e., it is a part of a greater system.

Here, a robotic dynamics is considered as an example by Z. R. Novakovic [3],

\begin{equation}  %(1)
\begin{array}{lll}
M(q)\ddot{q}+d(q,\dot{q})=u,
\end{array}
\end{equation}
where $q,u, d(q, \dot{q})\in R^{n}$ denote the joint coordinates vector, control vector, and the vector grouping the Coriolis centrifugal and gravitational forces or external disturbance, respectively. $M(q)=M(q)^{T}\in R^{n\times n}$ is the positive definite non-singular inertia matrix. By the computing torque technique, one can design a state feedback law
\begin{equation}  %(2)
\begin{array}{lll}
u=M(q)b+d(q,\dot{q}),
\end{array}
\end{equation}
where $b\in R^{n}$ is to be designed. Using the information about the joint-coordinates error $e=q_{d}-q$ ($q_{d}$ is the desired motion of the joints, assuming that the inverse kinematics problem has been solved), let
\[
b=\ddot{q}_{d}+K_{d}\dot{e}+K_{p}e,
\]
which guarantees that system (1) behaves according to
\begin{equation}  %(3)
\begin{array}{lll}
\ddot{e}+K_{d}\dot{e}+K_{p}e=0,
\end{array}
\end{equation}
where $K_{d},K_{p}$ are diagonal matrices whose elements are selected so to guarantee $e\rightarrow 0$ in advance. But practically, to consider the
issues of robustness to parameter uncertainties, external disturbances, sensor noise and computational complexity, etc., the controller (2) can not be obtained directly. To overcome this difficulty, the author considered
\begin{equation}  %(4)
\begin{array}{lll}
u=\hat{M}(\tilde{q})b+\hat{d}(\tilde{q},\dot{\tilde{q}}),
\end{array}
\end{equation}
where $\tilde{q}, \dot{\tilde{q}}$ are available measured values, $\hat{M}(\tilde{q}), \hat{d}(\tilde{q},\dot{\tilde{q}})$ are the estimated values of $M(q)$ and
$d(q,\dot{q})$ in model (1).
From the basic idea of the PSS,  essentially, the controller is seen as a part of (1), which means $u$ can also be substituted into the error system  by $M(q)\ddot{q}+d(q,\dot{q})$ to cancel some uncertain terms,
thus its maximal limitation $u_{\max}$ can be assumed to be estimated by the bound of $|M(q_{d})\ddot{q}_{d}|+|d(q,\dot{q})|$,
then the author proposed some practical tracking control algorithms based on the principle of self-support.

Under the basic idea of PSS, it is not necessary to know the accurate values of $q, q_{d}$, and only the estimated error information is enough to
design $u$ such that $e$ can be driven into a fixed neighborhood of zero $D_{\varepsilon}$.
   For simplicity, let $q=[q_{1}, ...,q_{n}]^{T}$, $u=[u_{1}, ..., u_{n}]^{T}$, $b=diag(b_{i}), (i=1,2,...,n)$, $q_{d}=[q_{1d},..., q_{nd}]^{T}$, tracking error
$e=[e_{1},...,e_{n}]^{T}$ with $e_{i}=q_{id}-q_{i}$, $\max\{|u_{i}|\}=u_{i\max}$.

When estimating $q$ by $\tilde{q}$, we suppose $\tilde{e}_{i}=e_{i}-\int_{0}^{t}\omega_{i}(t)dt$ for all $t$, where
$\omega_{i}(t)$ is the measurement function, which is supposed to be bounded ($|\omega_{i}(t)|\leq c_{i1}$) and belong to a class of bounded integrable functions in the sense of Lebesgue integration, $i.e., \omega_{i}(t)\in L^{1}_{[0,t]}(f(t))\triangleq \big\{f(t):\int _{[0,t]}|f(s)|ds\leq c_{i2}\big\}$,
where $c_{i1}, c_{i2}$ are two positive constants given in advance.
A PSS feedback law is proposed by $u_{i}=-b_{i}s_{i}$, where $b_{i}>0$ is a design parameter to be given later,
$s_{i}=\dot{\tilde{e}}_{i}+\rho_{i}\tilde{e}_{i}$, $\rho_{i}>0$. And next, for given a small positive constant $\varepsilon$, we will state that the tracking error $e_{i}(t)$ can be driven into the neighborhood of zero $D_{\varepsilon}\triangleq \big\{e_{i}: |e_{i}|\leq \frac{\rho_{i}c_{i2}+c_{i1}}{\rho_{i}}+\varepsilon\big\}$ by selecting proper design parameters $b_{i}$.

To show how to select the design parameter $b_{i}$,
take a Lyapunov function $V_{1}=\frac{1}{2}\sum_{i=1}^{n}e_{i}^{2}$, its time derivative can be calculated
\[\hspace{-40mm}
\dot{V}_{1}=\sum_{i=1}^{n}e_{i}(\dot{\tilde{e}}_{i}+\omega_{i}(t))
\]
\[\hspace{-23mm}
=\sum_{i=1}^{n}e_{i}(s_{i}-\rho_{i}\tilde{e}_{i}+\omega_{i}(t))
\]
\[\hspace{12mm}
=\sum_{i=1}^{n}e_{i}\Big(-\frac{u_{i}}{b_{i}}-\rho_{i}(e_{i}-\int_{0}^{t}\omega_{i}(t)dt)+\omega_{i}(t)\Big)
\]
\[\hspace{20mm}
=-\sum_{i=1}^{n}\rho_{i}e_{i}^{2}-\sum_{i=1}^{n}e_{i}\Big(\frac{u_{i}}{b_{i}}-\rho_{i}\int_{0}^{t}\omega_{i}(t)dt+\omega_{i}(t)\Big),
\]
under the boundedness conditions of $u_{i}$, $\omega_{i}(t)$ and $\int_{0}^{t}\omega_{i}(t)dt$,  one has

\[
\dot{V}_{1}\leq -\sum_{i=1}^{n}\rho_{i}e_{i}^{2}+\sum_{i=1}^{n}|e_{i}|(\frac{u_{i\max}}{b_{i}}+\rho_{i}c_{i2}+c_{i1}),
\]
from which, if $|e_{i}|>\frac{\rho_{i}c_{i2}+c_{i1}}{\rho_{i}}+\varepsilon$, we have
\[
\dot{V}_{1}\leq -\sum_{i=1}^{n}\rho_{i}(\frac{\rho_{i}c_{i2}+c_{i1}}{\rho_{i}}+\varepsilon)|e_{i}|+\sum_{i=1}^{n}|e_{i}|(\frac{u_{i\max}}{b_{i}}+\rho_{i}c_{i2}+c_{i1})
\]
\[\hspace{-54mm}
=-\sum_{i=1}^{n}|e_{i}|(\rho_{i}\varepsilon-\frac{u_{i\max}}{b_{i}}).
\]
We can select design parameters $b_{i}$ such that $\eta=\rho_{i}\varepsilon-\frac{u_{i\max}}{b_{i}}>0$,  so
choosing $b_{i}>\frac{u_{i\max}}{\rho_{i}\varepsilon}$ such that
\[
\dot{V}_{1}\leq -\eta \sum_{i=1}^{n}|e_{i}| \leq 0,
\tag{*}
\]
which means $e_{i}(t)$ will enter into the region $D_{\varepsilon}$ in a finite time.

On the other hand, once $e_{i}(t) \in D_{\varepsilon}$, it has $|e_{i}(t)|\leq \frac{\rho_{i}c_{i2}+c_{i1}}{\rho_{i}}+\varepsilon$
and $|\dot{e}_{i}(t)|$ is also shown to be upper bounded, since

\[
|\dot{e}_{i}|=|\dot{\widetilde{e}}_{i}+\omega_{i}(t)|=|s_{i}-\rho_{i}\widetilde{e}_{i}+\omega_{i}(t)|,
\]
substituting control law and estimated error, we have

\[\hspace{-34mm}
|\dot{e}_{i}|=|-\frac{u_{i}}{b_{i}}-\rho_{i}\widetilde{e}_{i}+\omega_{i}(t)|
\]
\[
=|-\frac{u_{i}}{b_{i}}-\rho_{i}(e_{i}-\int_{0}^{t}\omega_{i}(t)dt)+\omega_{i}(t)|
\]
\[\hspace{3mm}
\leq \frac{u_{i\max}}{b_{i}}+\rho_{i}(\frac{\rho_{i}c_{i2}+c_{i1}}{\rho_{i}}+\varepsilon+c_{i2})+c_{i1}
\]

\[\hspace{-15mm}
=\frac{u_{i\max}}{b_{i}}+2\rho_{i}c_{i2}+2c_{i1}+\rho_{i}\varepsilon,
\]
because $b_{i}>\frac{u_{i\max}}{\rho_{i}\varepsilon}$, therefore
\[
|\dot{e}_{i}|<2\rho_{i}(c_{i2}+c_{i1}/\rho_{i}+\varepsilon).
\tag{**}
\]

This means that the control algorithm guarantees that $e_{i}$ will lie in $D_{\varepsilon}$ if $c_{i2}=c_{i1}=0, \varepsilon\rightarrow 0^{+}$.

\emph{Remark 1:} In an ideal world, $c_{i2}=c_{i1}=0$ means the sensors for  measuring the tracking error of robotic systems are  accurate without any
disturbance or noise, i.e., the properties of the final neighborhood of zero $D_{\varepsilon}$ depend on the  accuracy of sensors.
Therefore, a more generalized case (for any given $c_{i1}, c_{i2}$) of tracking problem is discussed here based on the basic PSS idea. Moreover, our further consideration in the next will be
the case when the estimated error is assumed to be measured by some cumulative error measurement function with memorability decided by the previous control effect.

\emph{Remark 2:} Usually, the desired objects to be tracked are moving in a bounded feasible region (the size of which may be very large), for all initial
conditions, from (*) and (**), both $\widetilde{e}_{i}(t)$ and $e_{i}(t)$ will not escape to infinite before $e_{i}(t)$ enters into $D_{\varepsilon}$.

Additionally, there are some research results about PSS in control systems, let's do a brief review on it. In [4],
 Tan et al. discussed the precision motion control of a permanent magnet linear motor (PMLM) for applications which are inherently repetitive in terms of the motion trajectories, a feedback-feedforward control structure is proposed with a modest amount of modeling effort. An iterative learning controller (ILC) based on zero-phase filtering is applied as feedforward controller to the existing relay-tuned PID feedback controller to enhance the trajectory tracking performance by utilizing the experience gained from the repeated execution of the same operations. Considering inputs subjected to bounded constraints, Novakovic [5] proposed a practical tracking algorithm, the control law is accelerometer-free (or even tacho-free, also), robust to sensor
noise and allows the prespecification of the error decay rate, and is realistic from the engineering standpoint that can be implemented using current microprocessor technology. The PSS methodology is introduced for kinematic control of manipulators, in which way that is both mathematically clear and simple
to implement [6]. Ulu et al. [7] proposed a  new method which is computationally more efficient, more suitable for coupling gain calculations of
arbitrary nonlinear contour and easier to implement on multiaxis systems, tracking and contouring performance of the method on a nonlinear contour is verified
through simulations and experiments achieving nanometer level accuracy for the two-axis system.

However, for complicated systems in engineering, designing an integer-order state feedback control law is imperfect especially when dealing with
some real-world plants need the so-called ``long term memory property" [8-9]. Compared with integer-order system, fractional calculus has been proven to describe real systems in interdisciplinary fields more effectively, since it can offer a deeper insight into the physical processes underlying a long-range memory behavior [10-14]. To sum up, fractional control related issues can include the fractional order dynamic system or plant to be controlled and the fractional-order controller. However, in control practice it is more common to consider the fractional-order controller [15]. This is due to the fact that the plant model may have already been obtained as an integer order model in the classical sense. In most cases, the task is to apply fractional-order control (FOC) to enhance the system control performance. For example, in [16], the robust control of perturbed integer-order LTI systems is considered by using a fractional order sliding surface design method. A novel control strategy has been proposed, ensuring that the fractional-order (FO) sliding manifold is hit at an infinite sequence of time instants becoming denser as time grows. The closed-loop system is proved to be asymptotically robust with respect to a wide class of disturbances with the chattering free FO sliding mode control. To improve control performance or for dealing explicitly with the fractional order behavior of the plants, in [17-18], the authors adopted
a fractional order PID controller or the generalized $PI^{\lambda}D^{\mu}$ controller. So, naturally, in this paper, we consider to present a
 fractional-order generalized principle self-support (FOGPSS) control for practice. What would be happen if the PSS controller (4) is replaced by FOGPSS controller? What condition should be satisfied compared with (3), and how to establish a FOGPSS feedback law?

The structure of the article is as follows: Section II presents the FOGPSS statement and a prospect of some possible research interests.
Section III provides a simple application example and its simulations.  And finally, a conclusion is summarized in Section VI.

\section{Problem Formulation of FOGPSS}

Among many definitions of fractional operators [19], commonly used are  Riemann-Liouville (RL) and Caputo fractional order operators.
The following subsection will give some basic definitions and properties.

\subsection{Preliminaries of Fractional Calculus}

\emph{Definition 1 ([20-21]):} Given function $f(t)$ at time instant $t\geq 0$, Riemann-Louville fractional integral with order $\alpha>0$ is defined as

\[
I^{\alpha}f(t)=D^{-\alpha}(t)=\frac{1}{\Gamma(\alpha)}\int_{0}^{t}\frac{f(\tau)}{(t-\tau)^{1-\alpha}}d\tau,
\]
where $\Gamma(\cdot)$ is the Euler gamma function [21],

\[
\Gamma(s)=\int_{0}^{\infty}e^{-t}t^{s-1}dt, \ \  s\in C.
\]
The reduction formula of this function holds

\[
\Gamma(s+1)=s\Gamma(s), \ \Rightarrow  \Gamma(m+1)=m(m-1)!=m!,
\]
where $m\in Z^{+}=\{1,2,3,...\}$.

\emph{Definition 2 ([20-21]):} The Riemann-Louville fractional derivative of function $f(t)$ with order $\alpha>0$  is defined as follows:

\[
^{RL}D^{\alpha}f(t)=\frac{1}{\Gamma(m-\alpha)}\frac{d^{m}}{dt^{m}}\int_{0}^{t}\frac{f(\tau)}{(t-\tau)^{\alpha-m+1}}d\tau,
\]
where $m-1<\alpha\leq m$ and $m\in Z^{+}$, $\frac{d^{m}}{dt^{m}}f(t)$ denotes $m-$ order derivative of $f(t)$ under the common sense.

\emph{Definition 3 ([20-21]):} The Caputo derivative of fractional order $\alpha$ of a function $f(t)$ is described by

\[
^{C}D^{\alpha}f(t)=D^{-(m-\alpha)}\frac{d^{m}}{dt^{m}}f(t)=\frac{1}{\Gamma(m-\alpha)}\int_{0}^{t}\frac{f^{(m)}(\tau)}{(t-\tau)^{1+\alpha-m}}d\tau,
\]
where $m-1\leq \alpha<m \in Z^{+}$.

For the fractional-order operators, we have the following properties [20-22]:

\emph{Property 1:} $I^{\alpha}I^{\beta}f(t)=I^{\alpha+\beta}f(t), \  \alpha, \beta\geq 0$.

\emph{Property 2:} $^{RL}D^{\alpha}\big(^{RL}D^{-\alpha}(f(t))\big)=f(t)$.

\emph{Property 3:} $^{C}D^{-\alpha}D^{-\beta}f(t)=D^{-(\alpha+\beta)}f(t),\  \alpha, \beta\geq 0$.

\emph{Property 4:} $^{C}D^{-\alpha}D^{\alpha}f(t)=f(t)-\Sigma_{j=0}^{m-1}\frac{m-1}{i!}f^{(j)(0)}$.

\emph{Property 5:} $^{C}D^{\alpha}I^{\alpha}f(t)=f(t)$.\\

Next, we will propose the fractional-order generalized principle of self-support (FOGPSS).

\subsection{Conception of FOGPSS}

The fractional-order generalized principle of self-support (FOGPSS) is proposed for us to design a fractional tracking error signals based on ``self-support"  to replace the
general control law. For instance, in order to to improve the control performance of robot dynamics (1), we consider to present a fractional-order error
state feedback in the PSS control law (4). This is not a simple replica of general PSS, but a challenging task both in control theory and in practical engineering application.

Since under the fractional PSS framework, the corresponding stability issue becomes the most urgent problem to solve, it's not clear that
the fractional-order asymptotic stability and Mittag-Leffler stability [23-28] can directly be applied to solve FOGPSS feedback design. In the
same example, if we propose a fractional state feedback with PSS in (4) , i.e., the undetermined term $b$ must satisfy some fractional-order
ordinary differential equation (ODE) corresponding to (3) such that the closed-loop error system will converge to a bounded neighborhood of zero
given in advance. In theory, this process will force the original system
into a pre-specified fractional-order error dynamics , it is a  big challenge  for practical plant with parametric or non-parametric uncertainty and nonlinearity
due to the imperfect stability criterion of nonlinear fractional-order systems.

Some useful stability theorems or conclusions  of fractional-order systems are listed as follows:

\emph{Lemma 1 ([29]):} For a differentiable vector $x(t)\in R^{n}$, and for any time instant $t\geq 0$,
\[
\frac{1}{2}\ ^{C}D^{\alpha}\big[x^{T}(t)x(t)\big]\leq x^{T}(t)^{C}D^{\alpha}x(t).
\]

\emph{Lemma 2 ([23]):} Let $^{C}D^{\alpha}x(t)\geq$$^{C}D^{\alpha}y(t)$, $\forall \alpha \in (0,1)$ and $x(0)=y(0)$,
then $x(t)\geq y(t)$.

\emph{Lemma 3 ([30]):} The linear fractional-order system with commensurate order $0<\alpha\leq 1$

\[
^{C}D^{\alpha}x(t)=Ax(t)
\]
is stable at $x=0$ if the following conditions are satisfied

\[
|arg(\lambda_{i})|>\alpha\frac{\pi}{2},
\]
where $\lambda_{i}$ are eigenvalues of matrix $A$.

\emph{Lemma 4([23]):} Consider the non-autonomous nonlinear fractional-order system

\begin{equation}  %(5)
\begin{array}{lll}
^{C}D^{\alpha}x(t)=f(x,t), \ \alpha\in (0,1),
\end{array}
\end{equation}
where $f: [0,\infty]\times \Omega\rightarrow R^{n}$ is piecewise continuous in $t$ and locally Lipschitz
in $x$ on $[0,\infty]\times \Omega$, and $\Omega\in R^{n}$ is a domain that contains an equilibrium point $x=0$.
If there exists a Lyapunov function $V(x(t),t)$ and class-K functions $\alpha_{i}, (i=1,2,3)$ satisfying

\[
\alpha_{1}(\|x\|)\leq V(x(t),t)\leq \alpha_{2}(\|x\|), \ ^{C}D^{\beta}V(x(t),t)\leq -\alpha_{3}(\|x\|),
\]
where $\beta\in (0,1)$.
Then the origin of system (5) is asymptotically stable.

On the other hand, to solve the FOGPSS, the available algorithms of fractional-order controller to be implemented in real time should be adopted.
Two approximation methods are the most frequently used to  calculate a linear or nonlinear fractional differential equation (FDE). One is
the Adams-Bashford-Moulton (ABM) algorithm, the other is the time-domain method which is a generalization of the ABM approximation algorithm.
This method is based on a predictor-corrector scheme using the Caputo definition [31]. We give a brief introduction of this algorithm as follows.

Consider the following fractional-order differential equation:
\begin{equation}  %(6)
D^{\alpha}x(t)=f(t,x(t)), \ 0\leq t\leq T
\end{equation}
with $x^{(k)}(0)=x_{0}^{(k)}$ $(k=0,1,2,..., \lceil\alpha\rceil-1)$.
Equation (6) is equivalent to the following Volterra integral equation
\begin{equation}  %(7)
x(t)=\sum_{k=0}^{\lceil\alpha\rceil-1}\frac{t^{k}}{k!}x_{0}^{(k)}
+\frac{1}{\Gamma(\alpha)}\int_{0}^{t}(t-\tau)^{\alpha-1}f(\tau,x(\tau))d\tau
\end{equation}

Set $h=T/N (N\in Z^{+})$, and $t_{n}=nh (n=0,1,2,...,N)$. Then (7) can be
discretized as follows:
\[
x_{h}(t_{n+1})=\sum_{k=0}^{\lceil\alpha\rceil-1}\frac{t_{n+1}^{k}}{k!}x_{0}^{(k)}
+\frac{h^{\alpha}}{\Gamma(\alpha+2)}f(t_{n+1},x_{h}^{p}(t_{n+1}))
+\frac{h^{\alpha}}{\Gamma(\alpha+2)}\sum_{j=0}^{n}a_{j, n+1}f(t_{j}, x_{h}(t_{j})),
\]
where
\[
x_{h}^{p}(t_{n+1})=\sum_{k=0}^{\lceil\alpha\rceil-1}\frac{t^{k}_{n+1}}{k!}x_{0}^{(k)}
+\frac{1 }{\Gamma(\alpha)}\sum_{j=0}^{n}b_{j, n+1}f(t_{j}, x_{h}(t_{j})),
\]
\[a_{j,n+1}=\left\{
\renewcommand\arraystretch{1.2}\begin{array}{lll}
n^{\alpha+1}-(n-\alpha)(n-j)^{\alpha+1}, j=0\\
(n-j+2)^{\alpha+1}+(n-j)^{\alpha+1}, j=0\\
-2(n-j+1)^{\alpha+1}, 1\leq j\leq n\\
1, \ j=n+1,
\end{array}
\right. \label{t1}
\]
and
\[
b_{j, n+1}=\frac{h^{\alpha}}{\alpha}((n-j+1)^{\alpha}-(n-j)^{\alpha}).
\]
The estimation error of this this technique is
\[
e=\max_{j=0,1,2,...,N} |x(t_{j})-x_{h}(t_{j})|=O(h^{p}),
\]
where $p=\min(2, 1+\alpha)$.

\subsection{Possible Research Framework of FOGPSS}

We will discuss possible research framework of FOGPSS in this subsection, which mainly includes four aspects:
$\lambda-$tracking control, tracking of time-delay system, saturated practical tracking and robotic system control.

\begin{itemize}
  \item \ \emph{$\lambda-$tracking control}
\end{itemize}

$\lambda-$stabilization or $\lambda-$tracking means that the output cannot be controlled to a set-point
but into a $\lambda-$neighbourhood of the set-point (or the reference trajectory to be tracked), where
$\lambda>0$ is an arbitrarily small constant given in advance [32-33].
For a large class of multivariable nonlinear minimum-phase systems of relative degree one,
Allgower et al. [34] modified a known adaptive high-gain control strategy $u(t)=-k(t)y(t), \dot{k}(t)=\|y(t)\|^{2}$ to obtain
a $\lambda-$tracking in the presence of output corrupted noise. In [35], for a class of high-gain stabilizable multivariable linear infinite-dimensional systems,
an adaptive control law is proposed to achieve the approximate asymptotic tracking in the sense that the tracking error converges to a neighborhood
of zero with the arbitrary prescribed radius $\lambda>0$. And a sampled version of the high-gain adaptive $\lambda-$tracking controller is considered in [36], which
is motivated by sampling arises from the possibility that the output of a system may not be available continuously, but only at discrete time instants.
Recently, Ilchmann et al. [37-40] considered the temperature control for exothermic chemical reactors by $\lambda-$tracking approach with a feedback law
subjected to saturation constraints.

By the research motivation above, it is possible to consider the adaptive $\lambda-$tracking control under FOGPSS framework,  more specifically,
we design a error feedback controller

\[
u(t)=-k(t)e(t), \ e(t)=y(t)-y_{r}(t)
\]
where $y(t), y_{r}(t)$ are output and desired tracking reference signal, respectively. The control gain $k(t)$ satisfies a fractional-order $\lambda-$adaptive ODE

\begin{equation*}
^{C}D^{\alpha}(k(t))=\begin{cases} f(e(t), \lambda),&\text{$\|e(t)\|\geq \lambda$},\\
0,&\text{$\|e(t)\|< \lambda$},
\end{cases}\eqno{(14)}
\end{equation*}
where the function $f(e(t), \lambda)$ in (14) is to be designed such that $e(t)$ can be driven into a small $\lambda-$neighborhood of zero with pre-given
$\lambda$. The core task of FOGSS control is to find an eligible function $f$ so that the FO tracking error closed-loop system is asymptotically stable at zero.

\begin{itemize}
  \item \ \emph{Tracking of systems with time-delay}
\end{itemize}

Time delay is the property of a physical system by which the response to an applied force (action) is delayed in its effect [41-42].
Time delays are often encountered in many dynamic systems such as rolling mill systems, biological systems, metallurgical processing systems, network systems,
and so on [43-44]. It has been shown that the existence of time delays usually becomes the source of instability and degrading performance of systems [43].
Many researches have been devoted to the study of tracking control of systems with time-delay, for example, Fridman [45] considers the sampled-data control of linear systems under uncertain sampling with the known
upper bound on the sampling intervals, a time-dependent Lyapunov functional method in the developed framework of
input delay approach has been introduced for analysis of this linear system. For a class of perturbed
strict-feedback nonlinear time-delay systems, an adaptive fuzzy tracking control scheme has been presented by appropriately choosing Lyapunov-Krasovskii functionals
and hyperbolic tangent functions [46]. In [47], the robust tracking and model following for a class of linear systems with known multiple delayed state perturbations
, time-varying uncertain parameters, and disturbance has been considered. A class of continuous memoryless state feedback controllers for robust
tracking of dynamical signals are proposed, by which, the tracking error can be guaranteed to decreases asymptotically to zero.
By using separation technique and the norm of neural weight vector,
Wang et al. [48] presented a simple and effective control approach to address the tracking problem for non-affine pure-feedback system with multiple time-varying
delay states.  For nonlinear discrete-time Systems with time delays, the model reference output feedback fuzzy tracking control design
and optimal tracking control based on heuristic dynamic programming have been discussed in [49] and [50], respectively. The tracking control
for switched linear systems with time-delay is solved by using single Lyapunov function technique and a typical hysteresis switching law
so that the $H_{\infty}$ model reference tracking performance can be satisfied [51].
And Cho et al. [52] considerer the robustness problem in time-delay control in the presence of the nonlinear
friction dynamics of robot manipulators that is enhanced with a compensator based on internal model control.

Considering the following nonlinear dynamical system of the form [53-54] with input time delay

\begin{equation}  %(8)
\begin{array}{lll}
\dot{x}=Ax+B[f(x)+g(x)u(t-\tau)], \\
y=Cx, \\
\end{array}
\end{equation}
where

\[
A=
\left
[\begin{array}{ccccc} 0 & 1 & 0 & \cdots & 0 \\
 0 & 0 & 1 & \cdots & 0 \\
 \vdots & \vdots & \vdots & \vdots & \vdots \\
 0 & 0 & 0 & \cdots & 1 \\
 0 & 0 & 0 & 0 & 0 \\
\end{array}\right], \
B=
\left
[\begin{array}{c} 0  \\
 0 \\
 \vdots  \\
 0  \\
 1 \\
\end{array}\right], \
C^{T}=
\left
[\begin{array}{c} 1  \\
 0 \\
 \vdots  \\
 0  \\
 0 \\
\end{array}\right], \
\]
$x\in R^{n}$ is the state vector, $y, u \in R$ are the output and control input, respectively. $\tau$ denotes the constant of time-delay. Let
$y_{r}$ be the reference signal,  $e(t)=y-y_{r}$ is the tracking error.

Then how to propose a fractional time-delay feedback controller for system (8) or other nonlinear systems with input time delay or state time-delay
is an important Pioneering research for the first time to the best of the authors' knowledge. This study will touch on the field of stability issue
about fractional-order time-delay systems combining with the PSS control strategy.

\begin{itemize}
  \item \ \emph{Practical tracking with input saturation}
\end{itemize}

From a practical point of view, it is important to design saturated controllers for any mechanical systems.
That is because any actuator always has a limitation of the physical control inputs (input saturation) [55-68],
while the control input signals are a function of the system states, large initial conditions or unmodeled disturbances
may cause the controller to exceed physical limitations [69], therefore, lots of saturated controllers design methods
have been proposed. H. Chen et al. considered the saturated stabilization or tracking of dynamic nonholonomic mobile robots [55-57] and
robust control for these robotic systems under a fixed camera feedback with input saturation [60-65], respectively.
For the systems with time delay, continuous or discrete, linear or nonlinear systems have also been studied by feedback law subject to
input saturation constraints in [67-69].  And Z. Lin et al. [58,70] have given a semi-global exponential stabilization control strategy including
state feedback law or output feedback type for both discrete-time
systems and continuous linear time-invariant systems subject to input saturation.
In [59], the robust stabilization of spacecraft in the presence of input saturation constraints, parametric uncertainty, and external
disturbances has been addressed by two globally stable control algorithms.
In [67], based on LMIs technique, the theory of the composite nonlinear feedback control method
has been  considered for robust tracking and model following
of linear systems with time varying delays and input saturation.
Recently, the saturated control for multi-agent systems has become a hot research topic, for
example, H. Su et al. [66] studied the observer-based leader-following
consensus of a linear multi-agent system on switching networks, in which the input of each agent is subject to saturation.
A low-gain output feedback strategy to design the new observer-based consensus algorithms, without requiring
any knowledge of the interaction network topology.
Also, the global consensus problem of discrete-time multi-agent systems with input
saturation constraints under fixed undirected topologies has been discussed in [71], in which,
two two special cases are considered, where the agent model is either neutrally stable
or a double integrator.

Commonly, the saturation function $Sat_{\varepsilon}(\cdot)$ is a monotonically increasing function whose
saturated level is less than $\varepsilon$, i.e., $|Sat_{\varepsilon}(\cdot)|\leq \varepsilon$.
Examples of such saturation functions, for instance [56],
\[
Sat_{\varepsilon}(\tilde{z})=\varepsilon tanh(\tilde{z}),\  Sat_{\varepsilon}(\tilde{z})= \frac{2\varepsilon}{\pi}\arctan(\tilde{z}), \
Sat_{\varepsilon}(\tilde{z})=
\left\{
\begin{array}
    {r@{\quad:\quad}l}
    \varepsilon & if |\tilde{z}|\geq \varepsilon,\\
    \tilde{z}  & otherwise. \\
    \end{array}
\right.
\]
The difficulty of saturating practical tracking feedback based on FOGPSS lies in the fact that we are short of theoretical support
because there are only a few results about the control of fractional-order systems with input saturation [72]. It is
necessary to find a new fractional-order system control technique to support this framework in the near future.

\begin{itemize}
  \item \ \emph{Robotic dynamics control}
\end{itemize}

There are many types of robot systems such as rigid robot manipulators [73-82],  humanoid robots [83-87], underwater robots [88-95],
space robots [96-98], wheeled mobile robots [99-105], pipe robots[106-108], and so on. Among which, studying of a class of robot systems
subject to nonholonomic motion constraints becomes a research hot point, and control of  such mobile robots have attracted considerable attention from the research
community because of their practical applications and the theoretical challenges created by the nonholonomic nature of the constraints on it [109-112].
It is because control such systems is full of practically engineering interesting and theoretically challenging,
just as reported by  Brockett [113], any nonholonomic system can not be stabilized to a point with pure smooth (or even continuous) state feedback control law.
In order to overcome this design difficulty, many ingenious feedback stabilization methods have been proposed such as discontinuous feedback control law [60-65],
time-varying feedback law [55-57], hybrid feedback law [114-115], and optimal feedback law [116-118], etc.

\begin{center}
\includegraphics [scale=0.7,trim=200 250 200 300]{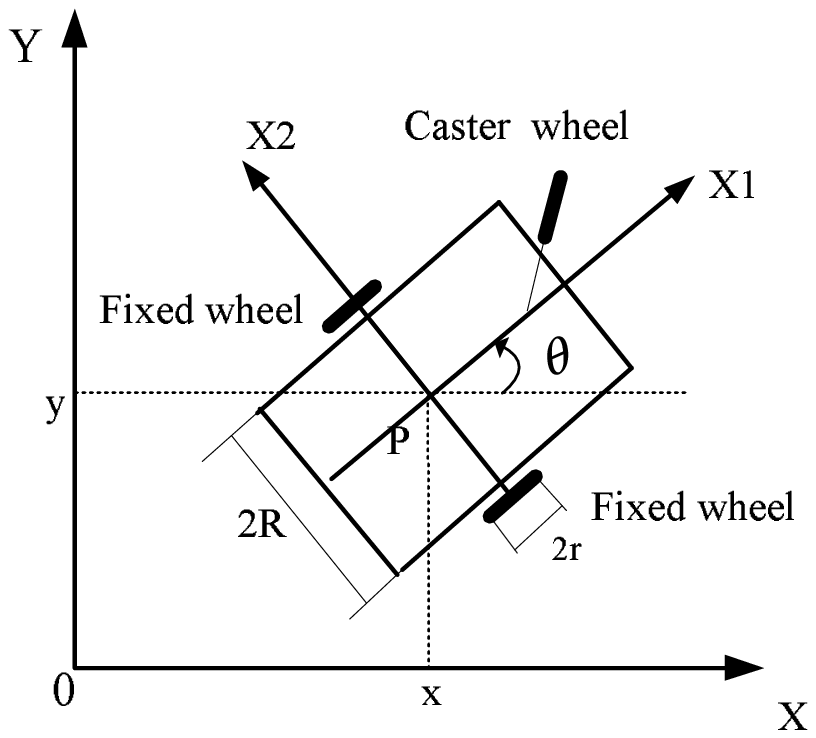}\\
\hspace{-13mm}\label{Fig1}{\fontsize{9.3pt}{11.6pt}\selectfont Fig.~4~~Nonholonomic wheeled mobile robot.}
\end{center}

As shown in Fig. 4, the  posture kinematic model of a class of nonholonomic wheeled mobile robots can be described by the following differential equations [99]:

\begin{equation}  %(9)
\left\{
\begin{array}{lll}
\dot{x}=v\cos\theta, \\
\dot{y}=v\sin\theta, \\
\dot{\theta}=\omega, \\
\end{array}
\right. \label{t1}
\end{equation}
where $(x,y)$ is the position of the mass center of the robot moving in the plane. $v$ is
the forward velocity,  $\omega$ is the steering velocity and $\theta$ denotes its heading angle from the horizontal axis.

Different from current approaches, FOGPSS tracking of the wheeled mobile robots (9) is independent of its desired trajectory $(x_{r}, y_{r},\theta_{r})$
with FO error state feedback $(x_{e}, y_{e}, \theta_{e})=(x-x_{r}, y-y_{r}, \theta-\theta_{r})$,

\[
\left\{
\begin{array}{lll}
\dot{x}_{r}=v_{r}\cos\theta_{r}, \\
\dot{y}_{r}=v_{r}\sin\theta_{r}, \\
\dot{\theta_{r}}=\omega_{r}. \\
\end{array}
\right. \label{t1}
\]
For the strong nonlinear robot system model (9), how to design some FO velocity controllers $(v, \omega)$ such that the error state $(x_{e}, y_{e},\theta_{e})$
converges to a small neighborhood of zero given in advance is an important future research objective.

\section{A Simple Application Example of FOGPSS}

\subsection {A simple tracking example}

A number of simple
systems of engineering interest may be represented by a first-order model, for
example, the braking of an automobile, the discharge of an electronic flash, or the
flow of fluid from a tank may be approximately represented by a first-order
differential equation [119]:
\begin{equation}  %(10)
\begin{array}{lll}
\dot{x}=-a_{p}x+b_{p}u+d(x,t),
\end{array}
\end{equation}
where $x,u\in R$ are the state and control input, respectively. $a_{p},b_{p}>0$ are bounded uncertain parameters (constants),
$d(x,t)$ is the external disturbance signal. Let $x_{d}(t)$ be a desired reference trajectory, $x_{e}=x_{d}-x$ is the tracking
error.

Here, the control objective is to present a FOGPSS feedback law $u$ such that error state $x_{e}$ can be driven into a
specified $\varepsilon_{0}-$neighbourhood of zero $D_{\varepsilon_{0}}$ with small positive constant $\varepsilon_{0}>0$ given in advance.

For practice, we make the following assumptions:

\emph{Assumption 1:} The position of $x_{d}$ to be tracked is not directly available, but it moves within a known bounded region with a constrained velocity,
i.e., $|x_{d}|\leq b_{1}, |\dot{x}_{d}|\leq b_{2}$, where $b_{1}, b_{2}>0$ are known constants.

\emph{Assumption 2:} There exists positive constants $\underline{a}, \bar{a}, \underline{b}, \bar{b}, \bar{d}$
for the follower system (10), such that for all $x$ and $t$,

\[
\underline{a}\leq |a_{p}|\leq \bar{a}, \ \underline{b}\leq |b_{p}|\leq \bar{b}, \ |d(x,t)|\leq \bar{d}.
\]

\emph{Assumption 3:} The estimate of error measurement $x_{e}$ can be denoted by

\[
\tilde{x}_{e}=x_{e}-I^{\alpha}\omega(t), \ \alpha\in (0,1)
\]
where the estimated error function $\omega(t)$ satisfies that

\[
|\omega(t)|\leq c_{1}, \  \ |I^{\alpha}\omega(t)|\leq c_{2}.
\]

By Assumption 2, note that the controller $u$ to be designed in (9) can be seen as an inherent part itself according to PSS [1,3,5-6],
that means

\begin{equation}  %(11)
|u|=\bigg|\frac{\dot{x}+a_{p}x-d(x,t)}{b_{p}}\bigg|\leq \frac{|\dot{x}|+\bar{a}|x|+\bar{d}}{\underline{b}}.
\end{equation}
Tracking the desired trajectory $x_{d}(t)$, and according to Assumption 1, it is entirely normal to suppose the
boundedness of $x, \dot{x}$ in some estimated, feasible motion region by $|x_{d}|$ and $|\dot{x}_{d}|$, hence, form (11), we assume that $|u|\leq u_{\max}$.

\emph{Remark 3:} Compared with the existing tracking problem, we suppose $x_{d}$ can not be obtained by designer directly in Assumption 1 is
more general. And therefore, in Assumption 3,  it is reasonable to assume there is an integrable error function $\omega(t)$ between $\tilde{x}_{e}$ and $x_{e}$
under the sense of fractional calculus (Def. 1) due to  the possible long term memory property in estimation of tracking error, because it is naturally to consider the current feedback relies on the previous tracking effects.

For being convenient, we denote $^{C}D^{\alpha}$ by $D^{\alpha}$, and the design results will be stated as follows:

\emph{Theorem 1:} Under Assumptions 1-3, for system (10), taking the FOGPSS feedback law

\begin{equation}  %(12)
u=\bar{\beta} \tilde{s},
\end{equation}
where $\bar{\beta}$ is a design parameter satisfies that
\[
\bar{\beta}>\frac{u_{\max}}{\delta\varepsilon_{0}}>0,
\]
where $\delta>0$ is also a design parameter, $\tilde{s}$ is the fractional-order estimated error feedback signal

\begin{equation}  %(12)
\tilde{s}=D^{\alpha}\tilde{x}_{e}+\delta\tilde{x}_{e}.
\end{equation}
Then the real tracking error $x_{e}$ will be driven into
$D_{\varepsilon_{0}}\triangleq \{x_{e}: |x_{e}|\leq \frac{\delta c_{2}+c_{1}}{\delta}+\varepsilon_{0}\} $.

\begin{proof}{}
Take a Lyapunov function $V=\frac{1}{2}x^{2}_{e}$, by applying Lemma 1, we have
\[
D^{\alpha}V=D^{\alpha}(\frac{1}{2}x^{2}_{e})\leq x_{e}D^{\alpha}x_{e},
\]
by Assumption 3 and Property 5 in Definition 3, it has
\[
D^{\alpha}V\leq x_{e}D^{\alpha}(\tilde{x}_{e}+I^{\alpha}\omega(t))=x_{e}(D^{\alpha}(\tilde{x}_{e})+\omega(t)).
\]
Substituting (13) into the formula above, it has
\[\hspace{-20mm}
D^{\alpha}V\leq x_{e}(\tilde{s}-\delta\tilde{x}_{e}+\omega(t))
\]
\[\hspace{-8mm}
=x_{e}(\frac{u}{\bar{\beta}}-\delta\tilde{x}_{e}+\omega(t))
\]
\[\hspace{13mm}
=x_{e}(\frac{u}{\bar{\beta}}-\delta(x_{e}-I^{\alpha}\omega(t))+\omega(t)).
\]
According to Assumption 3 again, we have
\begin{equation}  %(13)
D^{\alpha}V\leq -\delta x_{e}^{2}+|x_{e}|(\frac{u_{\max}}{\bar{\beta}}+\delta c_{2}+c_{1}).
\end{equation}

If $|x_{e}|>\frac{\delta c_{2}+c_{1}}{\delta}+\varepsilon_{0}$, from (14), we can obtain
\[
D^{\alpha}V\leq -\delta(\frac{\delta c_{2}+c_{1}}{\delta}+\varepsilon_{0})|x_{e}|+|x_{e}|(\frac{u_{\max}}{\bar{\beta}}+\delta c_{2}+c_{1})
\]
\[\hspace{5mm}
=-|x_{e}|(\delta(\frac{\delta c_{2}+c_{1}}{\delta}+\varepsilon_{0})-\frac{u_{\max}}{\bar{\beta}}-\delta c_{2}-c_{1})
\]
\[\hspace{-33mm}
=-\frac{|x_{e}|}{\bar{\beta}}(\bar{\beta}\delta\varepsilon_{0}-u_{\max}).
\]
Let $\hat{\beta}=\big(\bar{\beta}\delta\varepsilon_{0}-u_{\max}\big)/ \bar{\beta}$, from (12), since
$\bar{\beta}>\frac{u_{\max}}{\delta\varepsilon_{0}} >0$, so $\hat{\beta}>0$, which means
\[
D^{\alpha}V\leq -\hat{\beta}|x_{e}|=-\hat{\beta}\sqrt{2}V^{1/2}\leq 0,
\]
by Lemma 4, $x_{e}\rightarrow 0$ as $t\rightarrow \infty$, hence $x_{e}$ will be driven into $D\varepsilon _{0}$.

By the similar derivation process as (**) in the introduction section, once $|x_{e}|\leq \frac{\delta c_{2}+c_{1}}{\delta}+\varepsilon_{0}$, from (12)-(14), we
have

\[
|D^{\alpha}x_{e}|=|D^{\alpha}\widetilde{x}_{e}+\omega(t)|=|\frac{u}{\bar{\beta}}-\delta(x_{e}-I^{\alpha}\omega(t))+\omega(t)|
\leq 2\delta (c_{2}+c_{1}/\delta+\varepsilon_{0}).
\]
Then $|D^{\alpha}x_{e}|< \varepsilon_{0}\rightarrow 0^{+}$ as $c_{1}=c_{2}=0$, and according to Lemma 4,
this completes the proof.

\end{proof}

\subsection{Simulations}

In this subsection, when using FOGPSS tracking controller consists of (12) and (13), we adopt the approximate numerical ABM algorithm (6)-(7) for
solving the fractional differential equations for corresponding error  system of (10).

In the following simulations, given $\varepsilon_{0}=0.3$, we suppose that:
$\underline{a}=0.5$, $\bar{a}=1.5$, $\underline{b}=1.0$, $\bar{b}=2.0$, $d=0.5\sin(xt)$ and $\bar{d}=0.5$, $b_{1}=3.0$, $b_{2}=0.5$,
$c_{1}=0.1$, $c_{2}=1.5$, $\omega(t)=-0.045\cos(x_{e}t)$.
From (11), we can calculate that $u_{\max}=5.5$. According to Theorem 1, we choose design parameters: $\delta=10$, $\bar{\beta}=12$, $\alpha=0.3$, $\hat{\beta}=0.04$.
The initial conditions are $x(0)=-1.5$, $x_{d}(0)=0.5$.

\begin{center}
\includegraphics [scale=0.8,trim=200 15 200 5]{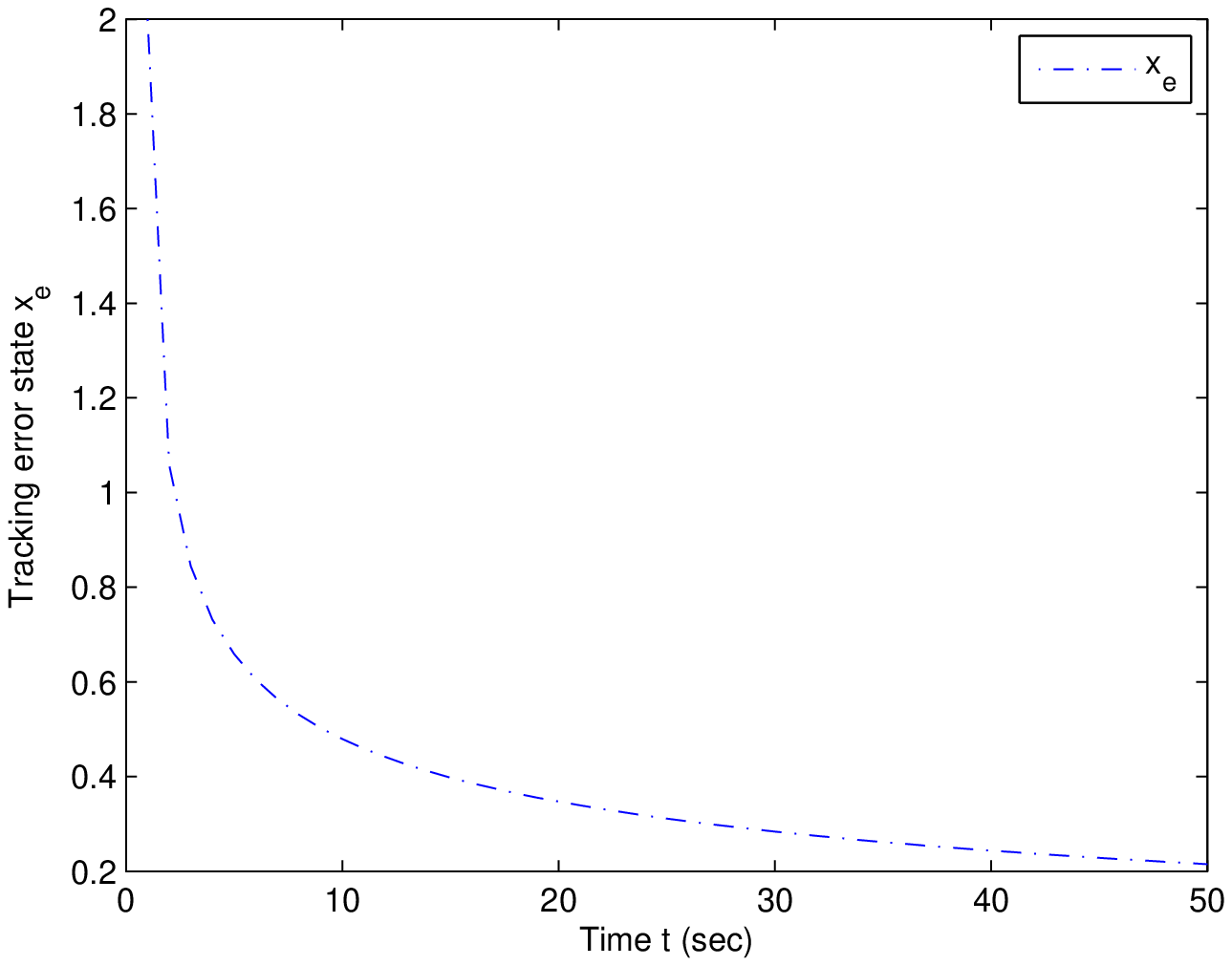}\\
\label{Fig4}{\fontsize{9.3pt}{11.6pt}\selectfont Fig.~5~~The response of tracking error state variable
$x_{e}$ with respect to time}
\end{center}

Some simulation results are shown in Figs. 5-7 with MATLAB. From Fig. 5, we can observe that the tracking error state $x_{e}$ is
driven into the small neighborhood of zero for given $\varepsilon_{0}=0.3$, surely, $|x_{e}|\leq 0.3$ at about $t\geq 30s$.
The response of estimated error state $\tilde{x}_{e}$ is demonstrated in Fig.6, from which, it can be seen that
the convergence behavior of $\tilde{x}_{e}$ is not like the $x_{e}$, since it is assumed that there exist a
error function $\omega(t)$ between $x_{e}$ and $\tilde{x}_{e}$, and $\tilde{x}_{e}$ goes into the $\varepsilon_{0}-$neighborhoos
of zero when $t\geq 20s$.
In Fig. 7, the  response of control input $u$ looks more like that of $\tilde{x}_{e}$ in Fig. 3 due to the FOGPSS feedback
consists of $D^{0.3}\tilde{x}_{e}$ and $\tilde{x}_{e}$ by (12) and (13).

\begin{center}
\includegraphics [scale=0.8,trim=200 0 200 10]{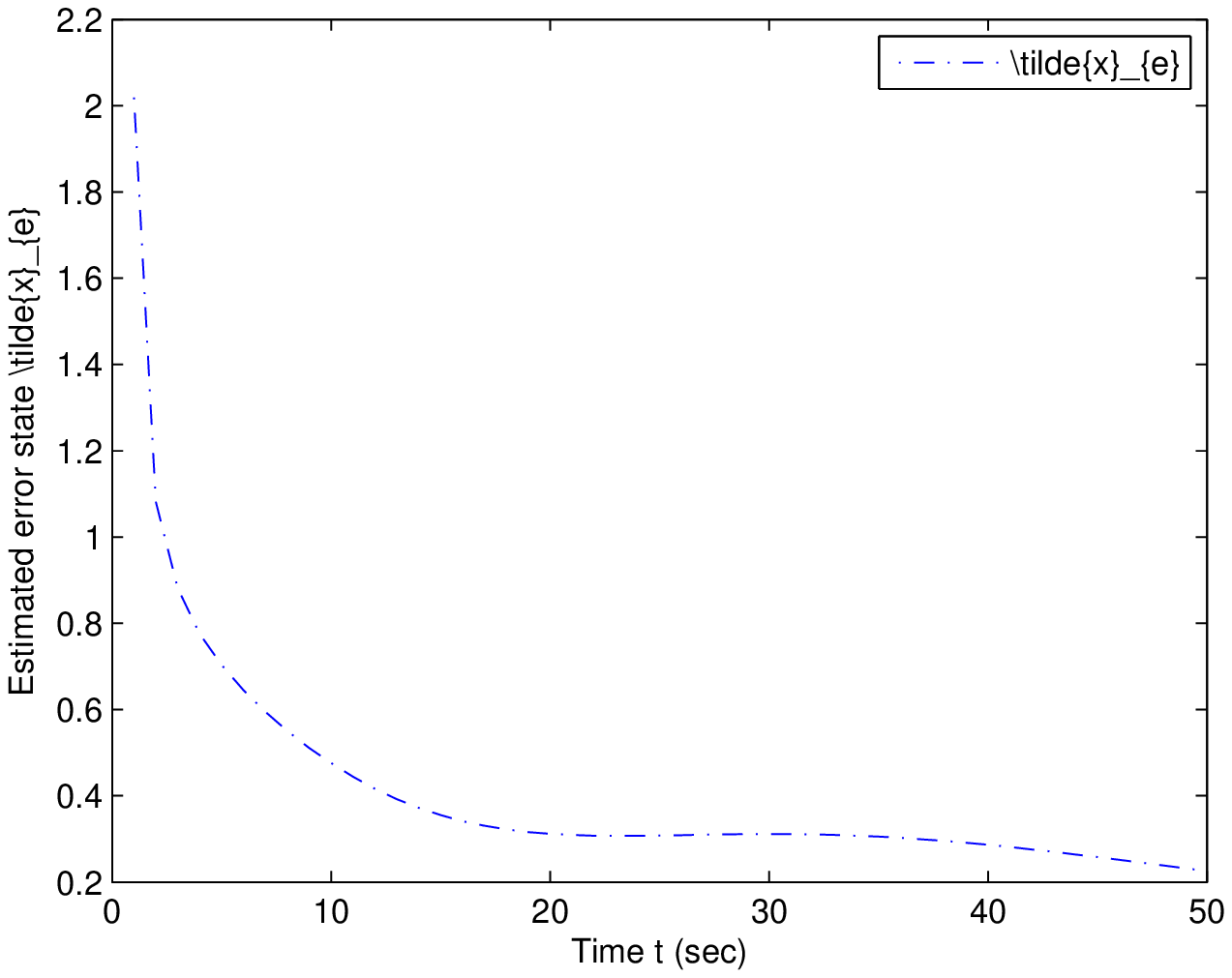}\\
\label{Fig5}{\fontsize{9.3pt}{11.6pt}\selectfont Fig.~6~~The response of estimated error state variable $\tilde{x}_{e}$ with respect to time}
\end{center}

\begin{center}
\includegraphics [scale=0.8,trim=200 0 200 0]{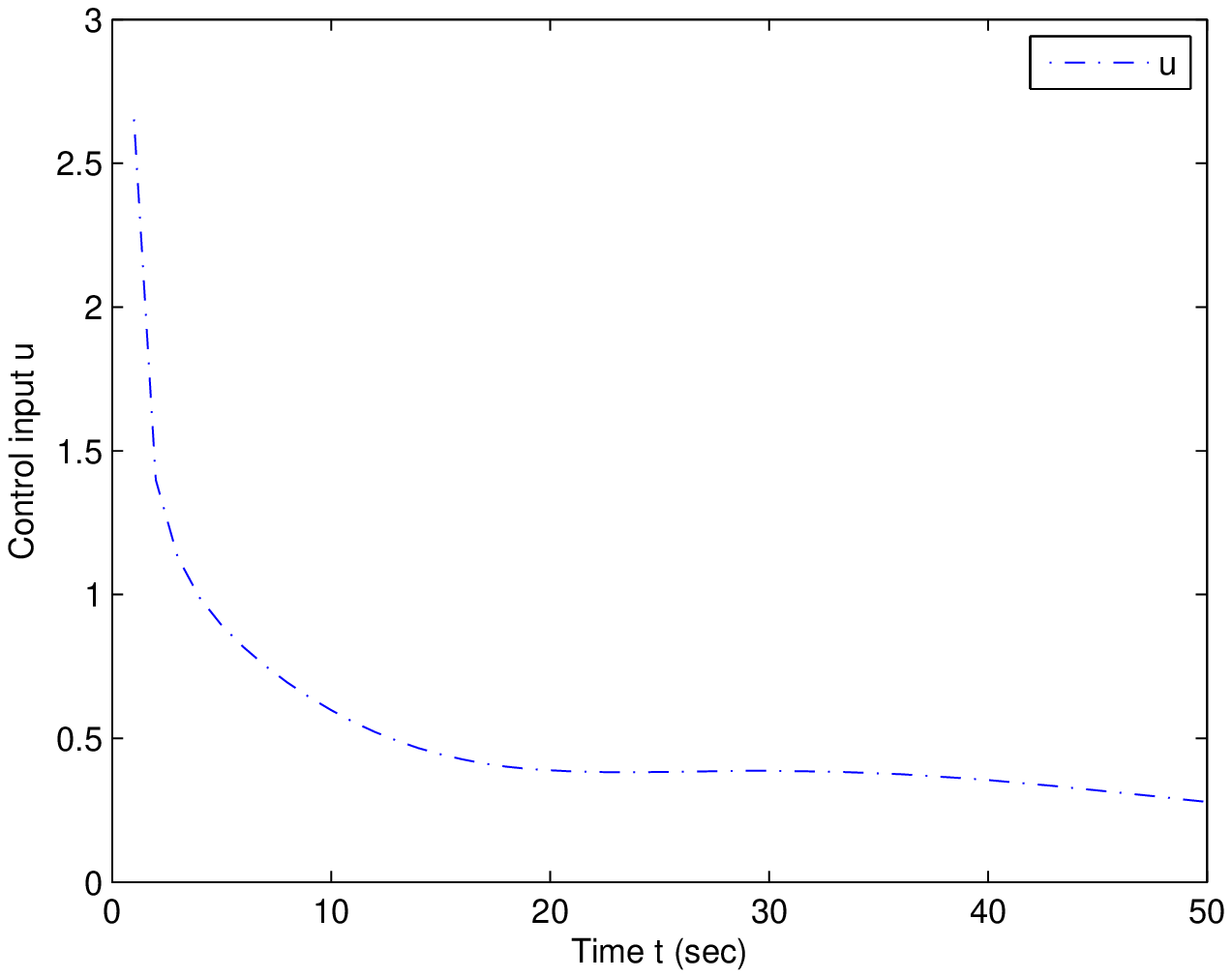}\\
\label{Fig6}{\fontsize{9.3pt}{11.6pt}\selectfont Fig.~7~~The response of FOGPSS state feedback input $u$ with respect to time}
\end{center}

\section{CONCLUSION}

In this article, a new conception of the generalized fractional-order principle of self-support (FOGPSS) is proposed for the first time.
After a brief reviews of PSS, the fractional-order-based framework is considered to  deal with the feedback control for practical complex system
,which is not perfect by integer-order feedback. And some possible research fields such as  practical tracking, $\lambda$-tracking, etc. for robot systems, multiple mobile agents, discrete dynamical systems, time delay systems
 and other uncertain nonlinear systems is discussed by FOGPSS. A simple example is presented to show the efficiency of the fractional-order  generalized principle of self-support (FOGPSS) control strategy.

 \section{ACKNOWLEDGEMENTS}

 This work was supported by the Natural Science Foundation of China (61304004, 61503205),
 the China Postdoctoral Science Foundation funded project (2013M531263),
 the Jiangsu Planned Projects for Postdoctoral Research Funds (1302140C),
 the Foundation of China Scholarship Council (201406715056),
 and the Foundation of Changzhou Key Laboratory of Special Robot and Intelligent Technology (CZSR2014005).

\end{spacing}

\end{document}